\documentclass{article}
\usepackage{graphicx}
\usepackage{amsmath}

\input{tcilatex}
\begin{document}

\noindent \noindent {\LARGE Calculation of the energy levels and sizes of
baryons with a noncentral harmonic potential}

\smallskip \medskip

\noindent M\'{a}rio Everaldo de Souza

\noindent \noindent Departamento de F\'{i}sica, Universidade Federal de
Sergipe

\noindent \noindent \noindent \noindent e-mail: mdesouza@sergipe.ufs.br
\smallskip

\noindent \medskip


\bigskip \medskip

{\small 
\parbox{4.5in} 
{Abstract. It is considered that the effective interaction between any two 
quarks in a baryon can be approximately described by a simple harmonic 
potential. Also, it is made use of the nonrelativistic approximation since 
the effective (constituent) masses of quarks are not small. The problem is 
firstly solved in Cartesian coordinates in order to find the energy levels 
irrespective of their angular momenta and then it is also solved in polar 
cilindrical coordinates for taking into account the angular momenta of the 
levels. By making a comparison between the two solutions the energies and 
the corresponding angular momenta (and parity) of almost all baryon levels 
are described. The agreement with the experimental data is quite impressive. 
The solution in Cartesian coordinates also produces some very important 
figures for the sizes of baryons and for the harmonic oscillator constant 
which is clearly related to confinement.} }

\section{\protect\normalsize \noindent Introduction}

{\normalsize \noindent }As is well known there are several important works
that deal with the calculation of the energy levels of baryons. One of the
most important ones is the pioneering work of Gasiorowicz and Rosner [1]
which has calculation of baryon levels and magnetic moments of baryons using
approximate wavefuncions. Another important work is that of Isgur and Karl
[2] which strongly suggests that non-relativistic quantum mechanics can be
used in the calculation of baryon spectra. Other very important attempts
towards the understanding of baryon spectra are the works of Capstick and
Isgur [3], Bhaduri et al.[4], Murthy et al.[5], \ Murthy et al.[6], and
Stassat et al.[7]. An important work attempting to describe baryon spectra
is the recent work of Hosaka, Toki and Takayama [8] published in 1998. This
last work arrives at an important equation which had already been deduced by
De Souza a long time ago, in 1992 [9]. Other works by De Souza published
before 1998 include it [10,11]$.$

\noindent \qquad The effective potential between any two quarks is not known
and because of this several different potentials are found in the
literature. In particular the harmonic approximation using a harmonic
central potential has been widely used. Since the three quarks of a baryon
are always in a plane it is assumed that the effective potential between any
two quarks of a baryon can be given by a linear harmonic oscillator. The
motion of the plane of quarks is not considered in this paper. This is a
calculation quite different from those found in the literature.

\section{Calculation in Cartesian coordinates}

\noindent In the initial calculation we use normal cartesian coordinates
which, of course, does not consider the angular momentum of the system, that
is, it does not take into account the symmetries of the system. But this
section is very important because it calculates the energy levels. In the
next section we will link each level to its angular momentum. Considering
the work of Isgur and Karl [2] as to the use of non-relativistic quantum
mechanics and using a linear harmonic oscillator potential [8,9] we can
write the Hamiltonian in normal cartesian coordinates as 
\begin{equation}
\sum_{i=1}^{6}\frac{{\partial }^{2}\psi }{{\partial }{\xi }_{i}^{2}}+\frac{2%
}{{\hbar }^{2}}\left( E-\frac{1}{2}\sum_{i=1}^{6}{\omega }_{i}{{\xi _{i}}^{2}%
}\right) \psi =0
\end{equation}
\noindent where we have used the fact that the three quarks are always in a
plane. The above equation may be resolved into a sum of 6 equations 
\begin{equation}
\frac{{\partial }^{2}\psi }{{\partial }{\xi }_{i}^{2}}+\frac{2}{{\hbar }^{2}}%
\left( E_{i}-\frac{1}{2}\omega _{i}{\xi _{i}}^{2}\right) \psi =0,
\end{equation}
\noindent which is the equation of a single harmonic oscillator of potential
energy $\frac{1}{2}\omega _{i}\xi _{i}^{2}$ and unitary mass with $%
E=\sum_{i=1}^{6}E_{i}$.

The general solution is a superposition of 6 harmonic motions in the 6
normal coordinates. The eigenfunctions $\psi _{i}(\xi _{i})$ are the
ordinary harmonic oscillator eigenfuntions 
\begin{equation}
\psi _{i}(\xi _{i})=N_{v_{i}}e^{-(\alpha _{i}/2)\xi _{i}^{2}}H_{v_{i}}(\sqrt{%
\alpha _{i}}\xi _{i}),
\end{equation}
\noindent where $N_{v_{i}}$ is a normalization constant, $\alpha _{i}=\nu
_{i}/\hbar $ and $H_{v_{i}}(\sqrt{\alpha _{i}}\xi _{i})$ is a Hermite
polynomial of the $v_{i}$th degree. For large $\xi _{i}$ the eigenfunctions
are governed by the exponential functions which make the eigenfunctions go
to zero very fast.

The energy of each harmonic oscillator is 
\begin{equation}
E_{i}=h\nu _{i}(v_{i}+\frac{1}{2}),
\end{equation}
\noindent where $v_{i}=0,1,2,3,...$ and $\nu _{i}$ is the classical
oscillation frequency of the normal ``vibration'' $i$, and $v_{i}$ is the
``vibrational'' quantum number. The total energy of the system can assume
only the values 
\begin{equation}
E(v_{1},v_{2},v_{3},...v_{6})=h\nu _{1}(v_{1}+\frac{1}{2})+h\nu _{2}(v_{2}+%
\frac{1}{2})+...h\nu _{6}(v_{6}+\frac{1}{2}).
\end{equation}

As was said above the three quarks in a baryon must always be in a plane.
Therefore, each quark is composed of two oscillators and so we may rearrange
the energy expression as 
\begin{equation}
E(n,m,k)=h\nu _{1}(n+1)+h\nu _{2}(m+1)+h\nu _{3}(k+1),
\end{equation}
\noindent where $n=v_{1}+v_{2},m=v_{3}+v_{4},k=v_{5}+v_{6}$. Of course, $%
n,m,k$ can assume the values, 0,1,2,3,... We may find the constants $h\nu $
from the ground states of some baryons. They are the known quark constituent
masses taken as $m_{u}=m_{d}=0.31$Gev, $m_{s}=0.5$Gev, $m_{c}=1.7$Gev,$%
m_{b}=5$Gev and $m_{t}=174$GeV.

\pagebreak

The states obtained with the above Hamiltonian are degenerate with respect
to isospin so that our calculation does not distinguish between nucleonic
and $\Delta $ states, or between $\Sigma $ and $\Lambda $ states. In the
tables below the experimental values of baryon masses were taken from
reference 12.

Let us start the calculation with the states ddu(neutron), uud(proton) and
ddd($\Delta ^{-}$), uuu ($\Delta ^{++}$) and their resonances. All the
energies below are given in Gev. Because $m_{u}=m_{d}$, we have that the
energies calculated by the formula 
\begin{equation}
E_{n,m,k}=0.31(n+m+k+3)
\end{equation}
\noindent correspond to many energy states. The calculated values are
displayed in Table 1. The last column on the right is a rough classification
which will be cleared up in the next section.

The energies of the particles $\Lambda $ and $\Sigma $, which are composed
of $uus$ and $uds$ are given by 
\begin{equation}
E_{n,m,k}=0.31(n+m+2)+0.5(k+1).
\end{equation}
\noindent The results are displayed in Table 2. The agreement with the
experimental values is excellent. \noindent For the $\Xi ^{o}$($uss$) and $%
\Xi ^{-}$($dss$) baryons the energies are expressed by 
\begin{equation}
E_{n,m,k}=0.31(n+1)+0.5(m+k+2).
\end{equation}
\noindent See Table 3 to check the agreement with the experimental data. In
this case the last column is almost empty due to a lack of experimental data.

In the same way the energies of $\Omega $($sss$) are obtained by 
\begin{equation}
E_{n,m,k}=0.5(n+m+k+3).
\end{equation}
\noindent The energies are displayed in Table 4. The discrepancies are
higher, of the order of 10\% and decreases as the energy increases. This is
a tendency which is also observed for the other particles. This may mean
that, at the bottom, the potential is less flat than the potential of a
harmonic oscillator.

The energies of the charmed baryons($C=+1$) $\Lambda _{c}^{+}$, $\Sigma
_{c}^{++}$, $\Sigma _{c}^{+}$ and $\Sigma _{c}^{0}$ are given by 
\begin{equation}
E_{n,m,k}=0.31(n+m+2)+1.7(k+1).
\end{equation}
\noindent The levels are shown in Table 5.

For the charmed baryons($C=+1$) $\Xi _{c}^{+}$ and $\Xi _{c}^{0}$ we have 
\begin{equation}
E_{n,m,k}=0.31(n+1)+0.5(m+1)+1.7(k+1).
\end{equation}
\noindent The results are displayed in Table 6.

As for the $\Omega _{c}^{0}$, its energies are 
\begin{equation}
E_{n,m,k}=0.5(n+m+2)+1.7(k+1).
\end{equation}
\noindent Table 7 shows the results of the energy levels.

In all tables below $E_{C}$ is the calculated value, $E_{M}$ is the measured
or experimental value and the Error is 
\[
Error=\left| \frac{E_{M}-E_{C}}{E_{C}}\right| \times 100\%. 
\]
As is easily seen in each table many levels may be predicted.\pagebreak

\begin{center}
\begin{tabular}{||c||c|l|c|l|c||}
\hline
&  &  &  &  &  \\ 
$n,m,k$ & $E_{C}(Gev)$ & $E_{M}$(Gev) & Error(\%) & $L_{2I.2J}$ & Parity \\ 
&  &  &  &  &  \\ \hline\hline
0,0,0 & 0.93 & 0.938($N$) & 0.9 & $P_{11}$ & + \\ \hline
$n+m+k=1$ & 1.24 & 1.232($\Delta $) & 0.6 & $P_{33}$ & + \\ \hline
$n+m+k=2$ & 1.55 & 1.52($N$) & 1.9 & $D_{13}$ & - \\ 
$n+m+k=2$ & 1.55 & 1.535($N$) & 1.0 & $S_{11}$ & - \\ 
$n+m+k=2$ & 1.55 & 1.6($\Delta $) & 3.1 & $P_{33}$ & + \\ 
$n+m+k=2$ & 1.55 & 1.62($\Delta $) & 4.5 & $S_{31}$ & - \\ \hline
$n+m+k=3$ & 1.86 & 1.90($N$) & 2.2 & $P_{13}$ & + \\ 
$n+m+k=3$ & 1.86 & 1.90($\Delta $) & 2.2 & $S_{31}$ & - \\ 
$n+m+k=3$ & 1.86 & 1.905($\Delta $) & 2.4 & $F_{35}$ & + \\ 
$n+m+k=3$ & 1.86 & 1.91($\Delta $) & 2.7 & $P_{31}$ & + \\ 
$n+m+k=3$ & 1.86 & 1.92($\Delta $) & 3.2 & $P_{33}$ & + \\ \hline
$n+m+k=4$ & 2.17 & 2.08($N$) & 4.1 & $D_{13}$ & - \\ 
$n+m+k=4$ & 2.17 & 2.09($N$) & 3.7 & $S_{11}$ & - \\ 
$n+m+k=4$ & 2.17 & 2.10($N$) & 3.2 & $P_{11}$ & + \\ 
$n+m+k=4$ & 2.17 & 2.15($\Delta $) & 0.9 & $S_{31}$ & - \\ 
$n+m+k=4$ & 2.17 & 2.19($N$) & 0.9 & $G_{17}$ & - \\ 
$n+m+k=4$ & 2.17 & 2.20($N$) & 1.4 & $D_{15}$ & - \\ 
$n+m+k=4$ & 2.17 & 2.20($\Delta $) & 1.4 & $G_{37}$ & - \\ 
$n+m+k=4$ & 2.17 & 2.22($N$) & 2.3 & $H_{19}$ & + \\ 
$n+m+k=4$ & 2.17 & 2.225($N$) & 5.5 & $G_{19}$ & - \\ \hline
$n+m+k=5$ & 2.48 & 2.39($\Delta $) & 3.6 & $F_{37}$ & + \\ 
$n+m+k=5$ & 2.48 & 2.40($\Delta $) & 3.2 & $G_{39}$ & - \\ 
$n+m+k=5$ & 2.48 & 2.42($\Delta $) & 2.4 & $H_{3,11}$ & + \\ \hline
$n+m+k=6$ & 2.79 & 2.7($N$) & 3.2 & $K_{1,13}$ & + \\ 
$n+m+k=6$ & 2.79 & 2.75($\Delta $) & 1.4 & $I_{3,13}$ & - \\ \hline
$n+m+k=7$ & 3.10 & 3.100($N$) & 0 & $L_{1,15}$ & ? \\ \hline
$n+m+k=8$ & 3.21 & ? & ? & ? & ? \\ \hline
$n+m+k=9$ & 3.72 & ? & ? & ? & ? \\ \hline
$n+m+k=9$ & 4.03 & ? & ? & ? & ? \\ \hline
... & ... & ... & ... & ... &  \\ \hline\hline
\end{tabular}

\vskip .2in

\parbox{4.5in}
{Table 1. Baryon states $N$ and $\Delta$. The energies $E_{C}$ were 
calculated according to the formula $E_{n,m,k} = 0.31(n+m+k+3)$ in which
$n,m,k$ are integers. $E_{M}$ is the measured energy.  The error means the 
absolute value of $(E_{C} - E_{M})/E_{C}$. We are able, of  course, to 
predict the energies of many other resonances.}
\end{center}

\pagebreak

\begin{center}
\begin{tabular}{||c||c|l|c|c|c||}
\hline\hline
&  &  &  &  &  \\ 
State($n,m,k$) & $E_{C}(Gev)$ & $E_{M}$(Gev) & Error(\%) & $L_{2I,2J}$ & 
Parity \\ 
&  &  &  &  &  \\ \hline\hline
0,0,0 & 1.12 & 1.116($\Lambda $) & 0.4 & $P_{01}$ & + \\ 
0,0,0 & 1.12 & 1.193($\Sigma $) & 6.5 & $P_{11}$ & + \\ \hline
$n+m=1$, k=0 & 1.43 & 1.385($\Sigma $) & 3.2 & $P_{13}$ & + \\ 
$n+m=1$, k=0 & 1.43 & 1.405($\Lambda $) & 1.7 & $S_{01}$ & - \\ 
$n+m=1$, k=0 & 1.43 & 1.48($\Sigma $) & 3.5 & ? & ? \\ \hline
0,0,1 & 1.62 & 1.52($\Lambda $) & 6.2 & $D_{03}$ & - \\ 
0,0,1 & 1.62 & 1.56($\Sigma $) & 3.7 & ? & ? \\ 
0,0,1 & 1.62 & 1.58($\Sigma $) & 2.5 & $D_{13}$ & - \\ 
0,0,1 & 1.62 & 1.60($\Lambda $) & 1.2 & $P_{01}$ & + \\ 
0,0,1 & 1.62 & 1.62($\Sigma $) & 0 & $S_{11}$ & - \\ 
0,0,1 & 1.62 & 1.66($\Sigma $) & 2.5 & $P_{11}$ & + \\ 
0,0,1 & 1.62 & 1.67($\Sigma $) & 3.1 & $D_{13}$ & - \\ 
0,0,1 & 1.62 & 1.67($\Lambda $) & 3.1 & $S_{01}$ & - \\ \hline
$n+m=2$, k=0 & 1.74 & 1.69($\Lambda $) & 2.9 & $D_{03}$ & - \\ 
$n+m=2$, k=0 & 1.74 & 1.69($\Sigma $) & 2.9 & ? & ? \\ 
$n+m=2$, k=0 & 1.74 & 1.75($\Sigma $) & 0.6 & $S_{11}$ & - \\ 
$n+m=2$, k=0 & 1.74 & 1.77($\Sigma $) & 1.7 & $P_{11}$ & + \\ 
$n+m=2$, k=0 & 1.74 & 1.775($\Sigma $) & 2.0 & $D_{15}$ & - \\ 
$n+m=2$, k=0 & 1.74 & 1.80($\Lambda $) & 3.4 & $S_{01}$ & - \\ 
$n+m=2$, k=0 & 1.74 & 1.81($\Lambda $) & 4.0 & $P_{01}$ & + \\ 
$n+m=2$, k=0 & 1.74 & 1.82($\Lambda $) & 4.6 & $F_{05}$ & + \\ 
$n+m=2$, k=0 & 1.74 & 1.83($\Lambda $) & 5.2 & $D_{05}$ & - \\ \hline\hline
\end{tabular}
\end{center}

\centerline{Continues on next page}

\pagebreak

\begin{center}
\begin{tabular}{||c||c|l|c|c|c||}
\hline\hline
&  &  &  &  &  \\ 
State($n,m,k$) & $E_{C}$(Gev) & $E_{M}$(Gev) & Error(\%) & $L_{2I,2J}$ & 
Parity \\ 
&  &  &  &  &  \\ \hline\hline
$n+m=1$, k=1 & 1.93 & 1.84($\Sigma $) & 4.7 & $P_{13}$ & + \\ 
$n+m=1$, k=1 & 1.93 & 1.88($\Sigma $) & 2.6 & $P_{11}$ & + \\ 
$n+m=1$, k=1 & 1.93 & 1.89($\Lambda $) & 2.1 & $P_{03}$ & + \\ 
$n+m=1$, k=1 & 1.93 & 1.915($\Sigma $) & 0.8 & $F_{15}$ & + \\ 
$n+m=1$, k=1 & 1.93 & 1.94($\Sigma $) & 0.5 & $D_{13}$ & - \\ \hline
$n+m=3$, k=0 & 2.05 & 2.00($\Lambda $) & 2.5 & ? &  \\ 
$n+m=3$, k=0 & 2.05 & 2.00($\Sigma $) & 2.4 & $S_{11}$ & - \\ 
$n+m=3$, k=0 & 2.05 & 2.02($\Lambda $) & 1.5 & $F_{07}$ & + \\ 
$n+m=3$, k=0 & 2.05 & 2.03($\Sigma $) & 1.0 & $F_{17}$ & + \\ 
$n+m=3$, k=0 & 2.05 & 2.07($\Sigma $) & 1.0 & $F_{15}$ & + \\ 
$n+m=3$, k=0 & 2.05 & 2.08($\Sigma $) & 1.5 & $P_{13}$ & + \\ \hline
0,0,2 & 2.12 & 2.10($\Sigma $) & 0.9 & $G_{17}$ & - \\ 
0,0,2 & 2.12 & 2.10($\Lambda $) & 0.9 & $G_{07}$ & - \\ 
0,0,2 & 2.12 & 2.11($\Lambda $) & 0.5 & $F_{05}$ & + \\ \hline
$n+m=2$, k=1 & 2.24 & 2.25($\Sigma $) & 0.5 & ? & ? \\ \hline
$n+m=4$, k=0 & 2.36 & 2.325($\Lambda $) & 1.5 & $D_{03}$ & - \\ 
$n+m=4$, k=0 & 2.36 & 2.35($\Lambda $) & 0.4 & $H_{09}$ & + \\ \hline
$n+m=1$, k=2 & 2.43 & 2.455 & 2.5 & ? &  \\ \hline
$n+m=3$, k=1 & 2.55 & 2.585($\Lambda $) & 1.4 & ? & ? \\ \hline
0,0,3 & 2.62 & 2.62($\Sigma $) & 0 & ? & ? \\ \hline
$n+m=5$, k=0 & 2.67 & to be found & ? & ? &  \\ \hline
$n+m=2$, k=2 & 2.74 & to be found & ? & ? &  \\ \hline
$n+m=4$, k=1 & 2.86 & to be found & ? & ? &  \\ \hline
$n+m=1$, k=3 & 2.93 & to be found & ? & ? &  \\ \hline
$n+m=6$, k=0 & 2.98 & 3.00($\Sigma $) & 0.7 & ? & ? \\ \hline
$n+m=3$, k=2 & 3.05 & to be found & ? & ? &  \\ \hline
$n=m=0$, k=4 & 3.12 & to be found & ? & ? &  \\ \hline
$n+m=5$, k=1 & 3.17 & 3.17($\Sigma $) & 0 & ? & ? \\ \hline
$n+m=2$, k=3 & 3.24 & to be found & ? & ? &  \\ \hline
& ... & ... & ... & ... & ... \\ \hline\hline
\end{tabular}
\end{center}

\vskip.2in

\begin{center}
\parbox{4.5in}
{Table 2. Baryon states $\Sigma$ and $\Lambda$. The energies $E_{C}$ were
calculated according to the formula $E_{n,m,k}= 0.31(n+m+2) + 0.5(k+1)$.
$E_{M}$ is the measured energy. The error  means the absolute value of 
$(E_{C} - E_{M})/E_{C}$.}
\end{center}

\pagebreak

\begin{center}
\begin{tabular}{||c||l|l|l|c|c||}
\hline
&  &  &  &  &  \\ 
State($n,m,k$) & $E_{C}$(Gev) & $E_{M}$(Gev) & Error(\%) & $L_{2I,2J}$ & 
Parity \\ 
&  &  &  &  &  \\ \hline\hline
0,0,0 & 1.31 & 1.315 & 0.5 & $P_{11}$ & + \\ \hline
1,0,0 & 1.62 & 1.53 & 5.6 & $P_{13}$ & + \\ 
1,0,0 & 1.62 & 1.62 & 0 & ? & ? \\ 
1,0,0 & 1.62 & 1.69 & 4.3 & ? & ? \\ \hline
n=0, $m+k=1$ & 1.81 & 1.82 & 0.6 & $D_{13}$ & - \\ \hline
2,0,0 & 1.93 & 1.95 & 1.0 & ? & ? \\ \hline
n=1, $m+k=1$ & 2.12 & 2.03 & 4.2 & ? & ? \\ 
n=1, $m+k=1$ & 2.12 & 2.12 & 0 & ? & ? \\ \hline
n=3, $m=k=0$ & 2.24 & 2.25 & 0.5 & ? & ? \\ \hline
n=0, $m+k=2$ & 2.31 & 2.37 & 2.6 & ? & ? \\ \hline
n=2, $m+k=1$ & 2.43 & to be found & ? & ? & ? \\ \hline
n=4, $m=k=0$ & 2.55 & 2.5 & 2.0 & ? & ? \\ \hline
n=1, $m+k=2$ & 2.62 & to be found & ? & ? & ? \\ 
& ... & ... & ... & ... & ... \\ \hline\hline
\end{tabular}
\end{center}

\vskip.2in

\begin{center}
\parbox{4.5in}
{Table 3. Baryon states $\Xi$. The energies $E_{C}$ were
calculated according to the formula $E_{n,m,k}= 0.31(n+1) + 0.5(m+k+2)$.
$E_{M}$ is the measured energy. The error means the absolute 
value of $(E_{C} - E_{M})/E_{C}$. The state $\Xi(1530)P_{13}$
appears to be the lowest state of the composite $\Xi\biguplus\pi$. Its
decay is in fact $\Xi\pi$.}
\end{center}

\vspace*{0.5in}

\begin{center}
\begin{tabular}{||c||l|c|l||}
\hline
&  &  &  \\ 
State($n,m,k$) & $E_{C}$(Gev) & $E_{M}$(Gev) & Error(\%) \\ 
&  &  &  \\ \hline\hline
0,0,0 & 1.5 & 1.672 & 11.7 \\ \hline
$n+m+k=1$ & 2.0 & 2.25 & 12.5 \\ \hline
$n+m+k=2$ & 2.5 & 2.47 & 1.2 \\ \hline
$n+m+k=3$ & 3.0 & to be found & ? \\ \hline
... & ... & ... & ... \\ \hline\hline
\end{tabular}
\end{center}

\vskip.2in

\begin{center}
\parbox{4.5in}
{Table 4. Baryon states $\Omega$. The energies $E_{C}$ were
calculated according to the formula $E_{n,m,k}= 0.5(n+m+k+3)$, and
$E_{M}$ is the measured energy.}
\end{center}

\vskip .3in

\vspace*{0.5in}

\begin{center}
\begin{tabular}{||c||l|c|l||}
\hline
&  &  &  \\ 
State($n,m,k$) & $E_{C}$(Gev) & $E_{M}$(Gev) & Error(\%) \\ 
&  &  &  \\ \hline\hline
0,0,0 & 2.32 & 2.285($\Lambda _{c}$) & 1.5 \\ \hline
$n+m=1$, k=0 & 2.63 & 2.594($\Lambda _{c}$) & 0.1 \\ 
$n+m=1$, k=0 & 2.63 & 2.627($\Lambda _{c}$) & 0.01 \\ \hline
$n+m=2$, k=0 & 2.94 & to be found & ? \\ \hline
... & ... & ... & ... \\ \hline\hline
\end{tabular}
\end{center}

\vskip.2in

\begin{center}
\parbox{4.5in}
{Table 5. Baryon states $\Lambda_{c}$ and $\Sigma_{c}$. The 
energies $E_{C}$ were calculated according to the formula 
$E_{n,m,k}= 0.31(n+m+2) + 1.7(k+1)$. The state with energy 2.63 MeV 
had already been predicted
in another version of this work. The experimental levels 2.594 MeV
and 2.627 MeV have confirmed the theoretical values. It appears that
the level $\Sigma_{c}(2.455)$ is a composition of the level 
$(0,0,0)$(that is the 2.285 $\Lambda_{c}$) with a pion as is also 
inferred from its decay.}
\end{center}

\vskip.5in

\begin{center}
\begin{tabular}{||c||l|c|l||}
\hline
&  &  &  \\ 
State($n,m,k$) & $E_{C}$(Gev) & $E_{M}$(Gev) & Error(\%) \\ 
&  &  &  \\ \hline\hline
0,0,0 & 2.51 & 2.47($\Xi _{c}^{+}$ & 1.6 \\ \cline{2-3}
& 2.51 & 2.47($\Xi _{c}^{0}$ & 1.6 \\ \hline
1,0,0 & 2.82 & to be found & ? \\ \hline
0,1,0 & 3.01 & to be found & ? \\ \hline
... & ... & ... & ... \\ \hline\hline
\end{tabular}
\end{center}

\vskip.2in

\begin{center}
\parbox{4.5in}
{Table 6. Baryon states $\Xi_{c}$. The energies $E_{C}$ were 
calculated according to the formula $E_{n,m,k}= 0.31(n+1) + 0.5(m+1) 
+ 1.7(k+1)$. $E_{M}$ is the measured energy. The recently found level 
$\Xi_{c}(2645)$ is probably a composition of the regular level 
$\Xi_{c}^{+}$ with a pion as its decay confirms.}
\end{center}

\pagebreak

\begin{center}
\begin{tabular}{||c||l|c|l||}
\hline
&  &  &  \\ 
State($n,m,k$) & $E_{C}$(Gev) & $E_{M}$(Gev) & Error(\%) \\ 
&  &  &  \\ \hline\hline
0,0,0 & 2.7 & 2.704($\Omega _{c}^{0}$ & 0 \\ \hline
$n+m=1$, k=0 & 3.2 & to be found & ? \\ \hline
$n+m=2$, k=0 & 3.7 & to be found & ? \\ \hline
... & ... & ... & ... \\ \hline\hline
\end{tabular}
\end{center}

\vskip.2in

\begin{center}
\parbox{4.5in}
{Table 7. Baryon states $\Omega_{c}$. The energies $E_{C}$ were 
calculated according to the formula $E_{n,m,k}= 0.5(n+m+2) + 1.7(k+1)$.
The energy of the level 
$(0,0,0)$ above shown had been predicted in other versions of this work.}
\end{center}

\vskip .3in

We may predict the energy levels of many other baryons given by the formulas:

\begin{itemize}
\item  ucc and dcc, $E_{n,m,k}=0.31(n+1)+1.7(m+k+2)$;

\item  scc, $E_{n,m,k}=0.5(n+1)+1.7(m+k+2)$;

\item  ccc, $E_{n,m,k}=1.7(n+m+k+3)$;

\item  ccb, $E_{n,m,k}=1.7(n+m+2)+5(k+1)$;

\item  cbb, $E_{n,m,k}=1.7(n+1)+5(m+k+2)$;

\item  ubb and dbb, $E_{n,m,k}=0.31(n+1)+5(m+k+2)$ ;

\item  uub, udb and ddb, $E_{n,m,k}=0.31(n+m+2)+5(k+1)$;

\item  bbb, $E_{n,m,k}=5(n+m+k+3)$;

\item  usb and dsb, $E_{n,m,k}=0.31(n+1)+0.5(m+1)+5(k+1)$;

\item  sbb, $E_{n,m,k}=0.5(n+1)+5(m+k+2)$;

\item  scb, $E_{n,m,k}=0.5(n+1)+1.7(m+1)+5(k+1)$;

\item  ucb, $E_{n,m,k}=0.31(n+1)+1.7(m+1)+5(k+1)$;

\item  ttt, $E_{n,m,k}=(174\pm 17)(n+m+k+3)$;

\item  and all combinations of t with u, d, c, s and b.
\end{itemize}

\section{Calculation in polar cilindrical coordinates}

In order to address the angular momentum and parity we have to use spherical
or polar coordinates. Since the three quarks of a baryon are always in a
plane we can use polar coordinates. We choose the Z axis perpendicular to
this plane. Now the eigenfunctions are angular momentum eigenfunctions (of
the orbital angular momentum). Thus, we have three oscillators in a plane.
Considering that they are independent the radial Schr\"{o}dinger equation
for the stationary states of each oscillator is given by [13] 
\begin{equation}
\left[ -\frac{\hbar ^{2}}{2\mu }\left( \frac{{\partial }^{2}}{{\partial \rho 
}^{2}}+\frac{1}{\rho }\frac{\partial }{\partial \rho }-\frac{m_{z}}{\rho ^{2}%
}\right) +\frac{1}{2}\mu \omega ^{2}\rho {^{2}}\right] R_{Em}(\rho
)=ER_{Em}(\rho )
\end{equation}
where $m_{z}$ is the quantum number associated to $L_{z}$. Therefore, what
we have is the following: three independent oscillators with orbital angular
momenta $L_{\mathbf{1}},\ L_{\mathbf{2}}$ and $L_{\mathbf{3}}$ which have
the Z components$\ L_{z1},L_{z2}$ and $L_{z3}$ in the plane containing the
quarks. Of course, the system has a total orbital angular momentum $L=$ $L_{%
\mathbf{1}}+\ L_{\mathbf{2}}+L_{\mathbf{3}}$ and there is a quantum number $%
l_{i}$ associated to each $L_{\mathbf{i}}$. The eigenvalues of the energy
are given by$^{13}$%
\begin{equation}
E=(2r_{1}+|m_{1}|+1)h\nu _{1}+(2r_{2}+|m_{2}|+1)h\nu
_{2}+(2r_{3}+|m_{3}|+1)h\nu _{3}
\end{equation}
in which $r_{1},r_{2},r_{3}=0,1,2,3,...$ and $|m_{i}|=0,1,2,3.....,l_{i}.$
Comparing the above equation with the equation 
\[
E(n,m,k)=h\nu _{1}(n+1)+h\nu _{2}(m+1)+h\nu _{3}(k+1), 
\]
we see that $n=2r_{1}+|m_{1}|$, $m=2r_{2}+|m_{2}|$, $k=2r_{3}+|m_{3}|$.

Let us recall that if we have three angular momenta $L_{1},L_{2}$ and $L_{3}$
described by the quantum numbers $l_{1},l_{2},l_{3}$ the total orbital
angular momentum $L$ will be described by the quantum number $l$ given by 
\begin{equation}
l_{1}+l_{2}+l_{3}\geq l\geq ||l_{1}-l_{2}|-l_{3}|
\end{equation}
where $l_{1}\geq |m_{1}|,$ $l_{2}\geq |m_{2}|,$ $l_{3}\geq |m_{3}|.$

Taking into account spin we form the total angular momentum given by $J=L+S$
and the quantum numbers of $J$ are $j=l\pm s$ where $s$ is the spin quantum
number. As we will see we will be able to describe almost all baryon levels.

\subsection{\noindent Baryons N and $\Delta $\ }

Let us begin the calculation with the particles $N$ and $\Delta $. We will
classify the levels by energy according to Table 1. The first state of $N$
is the state $(n=0,m=0,k=0)$ with energy 0.93 GeV. Therefore in this case $%
l_{1}=l_{2}=l_{3}=0$ and then $l=0$. Hence this is the positive parity state 
$P_{11}$ and we have

\smallskip

\begin{center}
{\normalsize 
\begin{tabular}{cccc}
\hline\hline
&  &  &  \\ 
$l$ & $N$ & $\Delta $ & Parity \\ 
&  &  &  \\ \hline
$0$ & $0.938P_{11}$ & $?$ & + \\ \hline\hline
\end{tabular}
}
\end{center}

\medskip

The second energy level (1.24\nolinebreak\ GeV) which is the first state of $%
\Delta $ has $n+m+k=1.$ This means that $%
2r_{1}+|m_{1}|+2r_{2}+|m_{2}|+2r_{3}+|m_{3}|=1$. Thus, $%
|m_{1}|+|m_{2}|+|m_{3}|=1$ and $l_{1}+l_{2}+l_{3}\geq 1$, and we can choose
the sets $%
|m_{1}|=1,|m_{2}|=|m_{3}|=0;|m_{1}|=|m_{3}|=0,|m_{2}|=1;|m_{1}|=1,|m_{2}|=|m_{3}|=0 
$, and $l_{1}=2,l_{2}=l_{3}=0$, or $l_{2}=2,l_{1}=l_{3}=0$, or still $%
l_{3}=2,l_{1}=l_{2}=0$ which produce $l=2$ and thus the level

\smallskip

\begin{center}
{\normalsize 
\begin{tabular}{cccc}
\hline\hline
&  &  &  \\ 
$l$ & $N$ & $\Delta $ & Parity \\ 
&  &  &  \\ \hline
$2$ & ? & 1.232$P_{33}$ & + \\ \hline\hline
\end{tabular}
}
\end{center}

\medskip

In the third energy level (1.55 Gev) $%
n+m+k=2=2r_{1}+|m_{1}|+2r_{2}+|m_{2}|+2r_{3}+|m_{3}|.$ This means that $%
|m_{1}|+|m_{2}|+|m_{3}|=2,0$ and we have the sets of possible values of $%
l_{1},l_{2},l_{3}$

\smallskip

\begin{center}
{\normalsize 
\begin{tabular}{c|ccccccc}
\hline\hline
$l_{1},l_{2},l_{3}$ & 2,0,0 & 0,2,0 & 0,0,2 & 1,1,0 & 1,0,1 & 0,1,1 & 0,0,0
\\ \hline
$l$ & 2 & 2 & 2 & 0,1,2 & 0,1,2 & 0,1,2 & 0 \\ \hline\hline
\end{tabular}
}
\end{center}

\smallskip

\noindent in which the second column presents the values of $l$ that satisfy
the condition $l_{1}+l_{2}+l_{3}\geq 2,0.$ There are thus the following
states

\smallskip

\begin{center}
{\normalsize 
\begin{tabular}{cccc}
\hline\hline
&  &  &  \\ 
$l$ & $N$ & $\Delta $ & Parity \\ 
&  &  &  \\ \hline
$0$ & $1.44P_{11}$, $1.71P_{11}$ & $?$ & + \\ \hline
$1$ & $
\begin{tabular}{c}
$1.535S_{11}$, $1.65S_{11}$ \\ 
$1.52D_{13}$, $1.70D_{13},1.675D_{15}$%
\end{tabular}
$ & $
\begin{tabular}{c}
$1.62S_{31}$ \\ 
$1.70D_{33}$%
\end{tabular}
$ & - \\ \hline
$2$ & $1.68F_{15},1.72P_{13}$ & $1.6P_{33}$ & + \\ \hline\hline
\end{tabular}
}
\end{center}

\smallskip


\noindent because we can have $%
j=1/2=0+1/2=1-1/2;j=3/2=1+1/2=2-1/2;j=5/2=1+3/2=2+1/2.$

The fourth energy level (1.86 Gev) has $%
n+m+k=3=2r_{1}+|m_{1}|+2r_{2}+|m_{2}|+2r_{3}+|m_{3}|$ which makes $%
|m_{1}|+|m_{2}|+|m_{3}|=3,1$ and $l_{1}+l_{2}+l_{3}\geq 3,1$. We have
therefore the possibilities

\smallskip

\begin{center}
\begin{tabular}{cc}
\hline\hline
$l_{1},l_{2},l_{3}$ & $l$ \\ \hline\hline
3,0,0 & 3 \\ \hline
0,3,0 & 3 \\ \hline
0,0,3 & 3 \\ \hline
2,1,0 & 3,2,1 \\ \hline\hline
\end{tabular}
\begin{tabular}{cc}
\hline\hline
$l_{1},l_{2},l_{3}$ & $l$ \\ \hline\hline
2,0,1 & 3,2,1 \\ \hline
1,0,2 & 3,2,1 \\ \hline
1,2,0 & 3,2,1 \\ \hline
0,1,2 & 3,2,1 \\ \hline\hline
\end{tabular}
\begin{tabular}{cc}
\hline\hline
$l_{1},l_{2},l_{3}$ & $l$ \\ \hline\hline
0,2,1 & 3,2,1 \\ \hline
1,0,0 & 1 \\ \hline
0,1,0 & 1 \\ \hline
0,0,1 & 1 \\ \hline\hline
\end{tabular}
\end{center}

\smallskip

\noindent and the states

\smallskip

\begin{center}
{\normalsize 
\begin{tabular}{cccc}
\hline\hline
&  &  &  \\ 
$l$ & $N$ & $\Delta $ & Parity \\ 
&  &  &  \\ \hline
$1$ & $2.08D_{13}$ & $
\begin{tabular}{c}
$1.90S_{31}$ \\ 
$1.94D_{33}$%
\end{tabular}
$ & - \\ \hline
$2$ & $1.90P_{13}$,$2.00F_{15}$, $1.99F_{17}$ & $1.91P_{31}$, $1.92P_{33}$, $%
1.905F_{35}$, & + \\ 
& $2.00F_{35}$,$1.95F_{37}$ & ? &  \\ \hline
$3$ & ? & $1.93D_{35}$ & - \\ \hline\hline
\end{tabular}
}
\end{center}

\medskip

In the fifth energy level (2.17 Gev) $n+m+k=4=$ $%
2r_{1}+|m_{1}|+2r_{2}+|m_{2}|+2r_{3}+|m_{3}|$ which yields $%
|m_{1}|+|m_{2}|+|m_{3}|=4,2,0$ and $l_{1}+l_{2}+l_{3}\geq 4,2,0$. We can
then have $l_{1}=l_{2}=l_{3}=0$ $(l=0)$ and also

\smallskip

\begin{center}
\begin{tabular}{cc}
\hline\hline
$l_{1},l_{2},l_{3}$ & $l$ \\ \hline\hline
4,0,0 & 4 \\ \hline
0,4,0 & 4 \\ \hline
0,0,4 & 4 \\ \hline
3,1,0 & 4,3,2 \\ \hline
3,0,1 & 4,3,2 \\ \hline
1,3,0 & 4,3,2 \\ \hline
1,0,3 & 4,3,2 \\ \hline
0,3,1 & 4,3,2 \\ \hline
0,1,3 & 4,3,2 \\ \hline\hline
\end{tabular}
\begin{tabular}{cc}
\hline\hline
$l_{1},l_{2},l_{3}$ & $l$ \\ \hline\hline
2,2,0 & 4,3,2,1,0 \\ \hline
2,0,2 & 4,3,2,1,0 \\ \hline
0,2,2 & 4,3,2,1,0 \\ \hline
2,0,0 & 2 \\ \hline
0,2,0 & 2 \\ \hline
0,0,2 & 2 \\ \hline
1,1,0 & 2,1,0 \\ \hline
1,0,1 & 2,1,0 \\ \hline
0,1,1 & 2,1,0 \\ \hline\hline
\end{tabular}
\end{center}

\smallskip

\noindent and hence the states

\smallskip


\begin{center}
{\normalsize 
\begin{tabular}{cccc}
\hline\hline
&  &  &  \\ 
$l$ & $N$ & $\Delta $ & Parity \\ 
&  &  &  \\ \hline
$0$ & $2.10P_{11}$ & ? & + \\ \hline
$1$ & $2.08S_{11}$, $2.20D_{15}$ & $2.15S_{31}$, $2.35D_{35}$ & - \\ \hline
$2$ & ? & $2.39F_{37}?$ & + \\ \hline
$3$ & $2.19G_{17}$, $2.25G_{19}$ & $2.20G_{37}$ & - \\ \hline
$4$ & $2.22H_{19}$ & $2.3H_{39}$ & + \\ \hline\hline
\end{tabular}
}
\end{center}

\medskip

In the sixth energy level (2.48 Gev) $%
n+m+k=5=2r_{1}+|m_{1}|+2r_{2}+|m_{2}|+2r_{3}+|m_{3}|$ which produces $%
|m_{1}|+|m_{2}|+|m_{3}|=5,3,1$ and $l_{1}+l_{2}+l_{3}\geq 5,3,1$. We have
then the possibilities

\smallskip

\begin{center}
\begin{tabular}{cc}
\hline\hline
$l_{1},l_{2},l_{3}$ & $l$ \\ \hline\hline
5,0,0 & 5 \\ \hline
0,5,0 & 5 \\ \hline
0,0,5 & 5 \\ \hline
4,1,0 & 5,4,3 \\ \hline
4,0,1 & 5,4,3 \\ \hline\hline
\end{tabular}
\begin{tabular}{cc}
\hline\hline
$l_{1},l_{2},l_{3}$ & $l$ \\ \hline\hline
1,4,0 & 5,4,3 \\ \hline
1,0,4 & 5,4,3 \\ \hline
0,4,1 & 5,4,3 \\ \hline
0,1,4 & 5,4,3 \\ \hline\hline
\end{tabular}
\begin{tabular}{cc}
\hline\hline
$l_{1},l_{2},l_{3}$ & $l$ \\ \hline\hline
3,0,0 & 3 \\ \hline
0,3,0 & 3 \\ \hline
0,0,3 & 3 \\ \hline
3,1,0 & 4,3,2 \\ \hline
3,0,1 & 4,3,2 \\ \hline\hline
\end{tabular}
\begin{tabular}{cc}
\hline\hline
$l_{1},l_{2},l_{3}$ & $l$ \\ \hline\hline
1,3,0 & 4,3,2 \\ \hline
1,0,3 & 4,3,2 \\ \hline
0,3,1 & 4,3,2 \\ \hline
0,1,3 & 4,3,2 \\ \hline\hline
\end{tabular}
\end{center}

\smallskip

\noindent Thus we identify the states

\smallskip

\begin{center}
{\normalsize 
\begin{tabular}{cccc}
\hline\hline
&  &  &  \\ 
$l$ & $N$ & $\Delta $ & Parity \\ 
&  &  &  \\ \hline
$2$ & ? & $2.39F_{37}$ & + \\ \hline
$3$ & ? & $2.40G_{39}$ & - \\ \hline
$4$ & ? & $2.42H_{3,11}$ & + \\ \hline
$5$ & $2.60I_{1,11}$ & ? & - \\ \hline\hline
\end{tabular}
}
\end{center}

\medskip

The seventh energy state (2.79 Gev) has $n+m+k=6$ $%
=2r_{1}+|m_{1}|+2r_{2}+|m_{2}|+2r_{3}+|m_{3}|$ which produces $%
|m_{1}|+|m_{2}|+|m_{3}|=6,4,2,0$ and $l_{1}+l_{2}+l_{3}\geq 6,4,2,0$. We
have then the possibilities\ below

\smallskip

\begin{center}
{\normalsize 
\begin{tabular}{c|ccccccc}
\hline\hline
$l_{1},l_{2},l_{3}$ & 6,0,0 & 0,6,0 & 0,0,6 & 5,1,0 & 5,0,1 & 1,5,0 & 1,0,5
\\ \hline
$l$ & 6 & 6 & 6 & 6,5,4 & 6,5,4 & 6,5,4 & 6,5,4 \\ \hline\hline
\end{tabular}
}
\end{center}

\smallskip

\noindent and the states

\smallskip


\begin{center}
{\normalsize 
\begin{tabular}{cccc}
\hline\hline
&  &  &  \\ 
$l$ & $N$ & $\Delta $ & Parity \\ 
&  &  &  \\ \hline
$4$ & ? & ? & + \\ \hline
$5$ & ? & $2.75I_{3,13}$ & - \\ \hline
$6$ & $2.7K_{1,13}$ & $2.95K_{3,15}$ & + \\ \hline\hline
\end{tabular}
}
\end{center}

\subsection{\noindent Baryons $\Sigma $\ and $\Lambda $}

Now let us do the calculation for $\Sigma $ and $\Lambda $. According to
Table 2 the first energy state (1.12 Gev) is $(n=0,m=0,k=0)$ and hence we
can have $l_{1}=0,l_{2}=0,l_{3}=0$ which yields $l=0$ and the states

\smallskip

\begin{center}
{\normalsize 
\begin{tabular}{cccc}
\hline\hline
&  &  &  \\ 
$l$ & $\Sigma $ & $\Lambda $ & Parity \\ 
&  &  &  \\ \hline
$0$ & $1.193P_{11}$ & $1.116P_{01}$ & + \\ \hline\hline
\end{tabular}
}
\end{center}

\medskip

In the second energy level (1.43 Gev) $n+m=1,k=0$ which makes $%
2r_{1}+|m_{1}|+2r_{2}+|m_{2}|=1$ and $2r_{3}+|m_{3}|=0$. This actually makes 
$|m_{1}|+|m_{2}|=1$ and $|m_{3}|=0.$ That is, we have the condition $%
l_{1}+l_{2}\geq 1,$ $l_{3}\geq 0$ which allows us to choose the possibilities

\smallskip

\begin{center}
{\normalsize 
\begin{tabular}{cccc}
\hline\hline
$l_{1},l_{2},l_{3}$ & 1,1,0 & 1,0,1 & 0,1,1 \\ \hline
$l$ & 2,1,0 & 2,1,0 & 2,1,0 \\ \hline\hline
\end{tabular}
}
\end{center}

\smallskip

\noindent that produce the states

\smallskip

\begin{center}
{\normalsize 
\begin{tabular}{cccc}
\hline\hline
&  &  &  \\ 
$l$ & $\Sigma $ & $\Lambda $ & Parity \\ 
&  &  &  \\ \hline
$0$ & $1.385P_{13}$ & ? & + \\ \hline
$1$ & ? & $1.405S_{01}$ & - \\ \hline
$2$ & ? & ? & + \\ \hline\hline
\end{tabular}
}
\end{center}

\smallskip

\noindent and the state $1.48\Sigma $ is either $S_{13},S_{11}(l=1)$ or $%
F_{15}(l=2).$


In the third energy level (1.62 Gev) $n=m=0,k=1$ and we have $%
|m_{1}|=0,|m_{2}|=0$ and $|m_{3}|=1.$ That is, we have the condition $%
l_{1}\geq 0,l_{2}\geq 0,$ $l_{3}\geq 1$ which allows us to choose $%
l_{1}=l_{2}=0,l_{3}=1;l_{1}=l_{3}=1,l_{3}=0$; $l_{1}=0,l_{2}=l_{3}=1$, and
the states

\smallskip

\begin{center}
{\normalsize 
\begin{tabular}{cccc}
\hline\hline
&  &  &  \\ 
$l$ & $\Sigma $ & $\Lambda $ & Parity \\ 
&  &  &  \\ \hline
$0$ & $1.66P_{11}$ & $1.60P_{01}$ & + \\ \hline
$1$ & $
\begin{tabular}{c}
$1.62S_{11}$ \\ 
$1.58D_{13}$%
\end{tabular}
$ & 
\begin{tabular}{c}
$1.67S_{01}$ \\ 
$1.52D_{03}$%
\end{tabular}
& - \\ \hline
$2$ & ? & ? & + \\ \hline\hline
\end{tabular}
}
\end{center}

\smallskip

\noindent and then the state $1.56\Sigma $ is probably $F_{15}($ $l=2)$.

The fourth energy level (1.74 Gev) has $n+m=2=2r_{1}+|m_{1}|+2r_{2}+|m_{2}|$
and $k=2r_{3}+|m_{3}|=0$, and thus we obtain $|m_{1}|+|m_{2}|=2,0$ and $%
|m_{3}|=0.$ Hence we have the condition $l_{1}+l_{2}\geq 2,0$ and $l_{3}\geq
0$. We can then choose $%
l_{1}=2,l_{2}=l_{3}=0;l_{1}=l_{3}=0,l_{2}=2;l_{1}=l_{2}=1,l_{3}=0$ and thus
we can identify the states

\smallskip

\begin{center}
\begin{tabular}{cccc}
\hline\hline
&  &  &  \\ 
$l$ & $\Sigma $ & $\Lambda $ & Parity \\ 
&  &  &  \\ \hline
$0$ & $1.77P_{11}$ & $1.81P_{01}$ & + \\ \hline
$1$ & $
\begin{tabular}{c}
$1.75S_{11}$ \\ 
$1.67D_{13},1.775D_{15}$%
\end{tabular}
$ & $
\begin{tabular}{c}
$1.80S_{01}$ \\ 
$1.69D_{03}$%
\end{tabular}
$ & - \\ \hline
$2$ & ? & $1.82F_{05}$ & + \\ \hline\hline
\end{tabular}
\end{center}

\smallskip

\noindent and then the level $1.69\Sigma $ is probably $F_{15}(l=2)$.

In the fifth energy level (1.93 Gev) $n+m=1=2r_{1}+|m_{1}|+2r_{2}+|m_{2}|$
and $k=1=2r_{3}+|m_{3}|$, and thus we obtain $|m_{1}|+|m_{2}|=1$ and $%
|m_{3}|=1.$ Hence we have the condition $l_{1}+l_{2}\geq 1$ and $l_{3}\geq 1$%
. We can then have the sets $l_{1}=1,l_{2}=0,l_{3}=1;$ $l_{1}=0,$ $%
l_{2}=1,l_{3}=1.$ Both yield $l=2,1,0$ and we can identify the states

\smallskip


\begin{center}
{\normalsize 
\begin{tabular}{cccc}
\hline\hline
&  &  &  \\ 
$l$ & $\Sigma $ & $\Lambda $ & Parity \\ 
&  &  &  \\ \hline
$0$ & $1.84P_{11},1.84P_{13}$ & $1.89P_{03}$ & + \\ \hline
$1$ & $1.94D_{13}$ & $1.83D_{05}$ & - \\ \hline
$2$ & $1.915F_{15}$ & $?$ & + \\ \hline\hline
\end{tabular}
}
\end{center}

\medskip

The sixth energy level (2.05 GeV) has $n+m=3=2r_{1}+|m_{1}|+2r_{2}+|m_{2}|$
and $k=0=2r_{3}+|m_{3}|$, and thus we obtain $|m_{1}|+|m_{2}|=3,1$ and $%
|m_{3}|=0.$ Hence we have the condition $l_{1}+l_{2}\geq 3,1$ and $l_{3}\geq
0$. We can then have the sets $l_{1}=2,l_{2}=1,l_{3}=0;$ $l_{1}=1,$ $%
l_{2}=2,l_{3}=0$ which make $l=3,2,1,$ for $l_{1}+l_{2}\geq 3$ and the sets $%
l_{1}=1,l_{2}=1,l_{3}=0;$ $l_{1}=1,$ $l_{2}=1,l_{3}=0$ which make $l=2,1,0,$
for $l_{1}+l_{2}\geq 1$. We can identify the states

\smallskip

\begin{center}
{\normalsize 
\begin{tabular}{cccc}
\hline\hline
&  &  &  \\ 
$l$ & $\Sigma $ & $\Lambda $ & Parity \\ 
&  &  &  \\ \hline
$0$ & $2.08P_{13}$ & ? & + \\ \hline
$1$ & $2.00S_{11}$ & ? & - \\ \hline
$2$ & $2.07F_{15},2.03F_{17}$ & $2.02F_{07}$ & + \\ \hline
$3$ & $?$ & ? & - \\ \hline\hline
\end{tabular}
}
\end{center}

\medskip

In the seventh energy level (2.12 GeV) $%
n=0=2r_{1}+|m_{1}|,m=0=2r_{2}+|m_{2}|,k=2=2r_{3}+|m_{3}|$ and thus $%
|m_{1}|=0,|m_{2}|=0,$ $|m_{3}|=2,0.$ Hence we have the condition $l_{1}\geq
0,l_{2}\geq 0$ and $l_{3}\geq 2,0$. We can then choose the sets $%
l_{1}=0,l_{2}=0,l_{3}=2;$ $l_{1}=0,$ $l_{2}=0,l_{3}=3$ which make $l=3,2$,
and the states

\smallskip

\begin{center}
{\normalsize 
\begin{tabular}{cccc}
\hline\hline
&  &  &  \\ 
$l$ & $\Sigma $ & $\Lambda $ & Parity \\ 
&  &  &  \\ \hline
$l=2$ & $?$ & $2.11F_{05}$ & + \\ \hline
$l=3$ & $2.10G_{17}$ & $2.10G_{07}$ & - \\ \hline\hline
\end{tabular}
}
\end{center}

\smallskip

Unfortunately, the angular momenta of the other energy levels have not been
found but they can surely be explained according to what was developed above.


\subsection{\noindent Baryons $\Xi ${\protect\normalsize \ }}

For these baryons only some angular momenta are known. The first energy
level (1.31 GeV) has $n=0,m=0.l=0$ which make $l_{1}=l_{2}=l_{3}=0$ and $l=0$
and is thus a $P$ state. Therefore we obtain

\smallskip

\begin{center}
{\normalsize 
\begin{tabular}{ccc}
\hline\hline
&  &  \\ 
$l$ & $\Xi $ & Parity \\ 
&  &  \\ \hline
$0$ & $1.318P_{11}$ & + \\ \hline\hline
\end{tabular}
}
\end{center}

\medskip

In the second energy level (1.62 GeV) $%
n=1=2r_{1}+|m_{1}|,m=0=2r_{2}+|m_{2}|,k=0=2r_{3}+|m_{3}|$ and thus $%
|m_{1}|=1,|m_{2}|=0,$ $|m_{3}|=0.$ Hence we have the condition $l_{1}\geq
1,l_{2}\geq 0$ and $l_{3}\geq 0$. We can then have the sets $%
l_{1}=1,l_{2}=0,l_{3}=0;$ $l_{1}=1,$ $l_{2}=1,l_{3}=0$ which make $l=2,1,0$,
and the states

\smallskip

\begin{center}
{\normalsize 
\begin{tabular}{ccc}
\hline\hline
&  &  \\ 
$l$ & $\Xi $ & Parity \\ 
&  &  \\ \hline
$0$ & $1.53P_{13}$ & + \\ \hline
$1$ & ? & - \\ \hline
$2$ & $?$ & + \\ \hline\hline
\end{tabular}
}
\end{center}

\smallskip

\noindent and thus the two levels $1.62$ and $1.69$ are probably either $S$, 
$D$ or $F$ states.

The third energy level (1.81GeV) has $%
n=0=2r_{1}+|m_{1}|,m+k=1=2r_{2}+|m_{2}|+2r_{3}+|m_{3}|$ and thus $%
|m_{1}|=0,|m_{2}|+$ $|m_{3}|=1.$ Hence we have the condition $l_{1}\geq
0,l_{2}+$ $l_{3}\geq 1$. We can then have the sets $l_{1}=0,l_{2}=1,l_{3}=0;$
$l_{1}=0,$ $l_{2}=0,l_{3}=1$ which make $l=1$, and the state

\smallskip

\begin{center}
{\normalsize 
\begin{tabular}{ccc}
\hline\hline
&  &  \\ 
$l$ & $\Xi $ & Parity \\ 
&  &  \\ \hline
$l=1$ & $1.82D_{13}$ & - \\ \hline\hline
\end{tabular}
}
\end{center}

\medskip

In the fourth energy level (1.93GeV) $%
n=2=2r_{1}+|m_{1}|,m=0=2r_{2}+|m_{2}|,k=0=2r_{3}+|m_{3}|$ and thus $%
|m_{1}|=2,0,|m_{2}|=0,$ $|m_{3}|=0.$ Hence we have the condition $l_{1}\geq
0,2,l_{2}\geq 0$ and $l_{3}\geq 0$. We can then choose the set $%
l_{1}=2,l_{2}=0,l_{3}=0$ which produces $l=2$, and the state 1.93GeV is
probably an $F$ state.

\subsection{\noindent Relation between energy and angular momentum}

From Eqs. 15 and 20 we have 
\[
\begin{array}{c}
E=(2r_{1}+|m_{1}|+1)h\nu _{1}+(2r_{2}+|m_{2}|+1)h\nu
_{2}+(2r_{3}+|m_{3}|+1)h\nu \\ 
l_{1}+l_{2}+l_{3}\geq l\geq ||l_{1}-l_{2}|-l_{3}|\text{ with }l_{1}\geq
|m_{1}|,l_{2}\geq |m_{2}|,l_{3}\geq |m_{3}|.
\end{array}
\]

\noindent in which $l_{1}$, $l_{2}$ and $l_{3}$ are the quantum numbers of
the angular momenta $\overrightarrow{L_{1}}$, $\overrightarrow{L_{2}}$, and $%
\overrightarrow{L_{3}}$, and $m_{1}$,$m_{2}$, $m_{3}$ are the quantum
numbers of their projections on the Z axis, respectively. Therefore, we
clearly see that levels with large energies have large angular momenta as is
quite evident from the experimental data.

\bigskip

\section{The sizes of baryons}

The solution in Cartesian coordinates is also useful for calculating in a
quite simple manner the average size of a baryon. As is known the average
potential energy of each oscillator is half of the total energy, that is,

\begin{equation}
<\frac{1}{2}k{{\xi _{i}}^{2}>=}\frac{h\nu _{i}}{2}(v_{i}+\frac{1}{2})
\end{equation}

\noindent but since there are two directions for each quark in the plane
there actually are two oscillators per quark and thus we have the potential
energy $E_{q}$ associated to each quark 
\begin{equation}
E_{q}=h\nu _{q}(n_{i}+1)
\end{equation}

\noindent where $n_{i}=0,1,2,3,...$ and $h\nu _{q}$ is the constituent quark
mass constant. Thus taking into account Eq. 06 and the above fact on the
relation between the total energy and the potential energy for an oscillator
it can be written that

\begin{equation}
\begin{array}{c}
E(n,m,k)=h\nu _{1}(n+1)+h\nu _{2}(m+1)+h\nu _{3}(k+1)= \\ 
=2\times \left( <\frac{1}{2}k_{1}\eta {{_{1}}^{2}>+}<\frac{1}{2}k_{2}\eta {{%
_{2}}^{2}>+}<\frac{1}{2}k_{3}\eta {{_{3}}^{2}>}\right) = \\ 
=\ <k_{1}\eta {{_{1}}^{2}>+}<k_{2}\eta {{_{2}}^{2}>+}<k_{3}\eta {{_{3}}^{2}>}
\end{array}
\end{equation}

\noindent where $\eta {{_{i}}^{2}=}$ ${{\xi _{ij}}^{2}+{\xi _{ik}}^{2}}$ in
which $j$ and $k$ are the two orthogonal directions of the two oscillators.
One can then make the association 
\begin{equation}
h\nu _{i}(n+1)=\,<k_{1}\eta {{_{1}}^{2}>}
\end{equation}

\noindent and hence the average radius $\mathcal{R\ }$\ of a baryon can
given by

\begin{equation}
\begin{array}{c}
\mathcal{R}(n,m,k)=\left( \sqrt{<\eta {{_{1}}^{2}>}<\eta {{_{2}}^{2}>}<\eta {%
{_{3}}^{2}>}}\right) ^{1/3}= \\ 
=\left( \sqrt{\frac{h\nu _{1}(n+1)}{k_{1}}\frac{h\nu _{2}(m+1)}{k_{2}}\frac{%
h\nu _{3}(k+1)}{k_{3}}}\right) ^{1/3}.
\end{array}
\end{equation}

It is quite obvious that the application of the above formula should be
first done to the proton. In the fundamental level $n=m=k=0$ and $h\nu
_{1}=h\nu _{2}=h\nu _{3}=$ $0.31$GeV, and making the reasonable supposition
that $k_{1}=k_{2}=k_{3}=k$, thus 
\begin{equation}
\mathcal{R}_{0}=\mathcal{R}(0,0,0)=\sqrt{\frac{h\nu _{1}}{k}}.
\end{equation}

\noindent If one uses for the average size of a proton the figure of $\sqrt{%
0.72}$fm = 0.85fm [14] one has $k\approx 0.5$GeV/fm$^{2}$ which is a very
reasonable figure because if it is multiplied by the characteristic distance
of 1fm (of course) the constant $k^{\prime }\approx 0.5$GeV/fm is obtained
which is quite close to the value of the constant $K$ used in the QCD
motivated potential [15], [16] 
\begin{equation}
V_{QCD}=-CF\frac{\alpha _{s}}{r}+Kr
\end{equation}
which is assumed to be of the order of 1GeV/fm.

From Table 1 one has that for $n=m=k=2$ the energy of a proton is about
2.80GeV which gives an average radius of about 1.39fm and hence one sees
that the size of a baryon does not change much with the its energy.
Therefore it can be said that the smallest radius of a proton is about 0.8fm
and the largest radius is or the order of 1.4fm.

For the ground states of $\ \Sigma ^{-}$ and $\Xi ^{-}$ reference [15]
gives, respectively, the radii $\sqrt{0.54}$fm = 0.73fm and $\sqrt{0.43}$fm
= 0.66fm. In terms of quarks $\Sigma ^{-}$ is $dus$ and therefore one should
have $k_{2}=k_{3}$ and $k_{1}\approx k\approx 0.5$GeV/fm$^{2}$ and 
\begin{equation}
\mathcal{R}(n,m,k)=\left( \sqrt{\frac{h\nu _{1}(n+1)\times h\nu
_{2}(m+1)\times h\nu _{3}(k+1)}{k_{1}(k_{3})^{2}}}\right) ^{1/3}.
\end{equation}
Using the above value it is obtained that 
\[
0.73=\left( \sqrt{\frac{0.31\times 0.31\times 0.5}{0.5\times (k_{3})^{2}}}%
\right) ^{1/3} 
\]

\noindent which yields $k_{3}=0.80$GeV/fm$^{2}$. From Table 2 it is seen
that for $n+m=5$, $k=1$, the energy is 3.17GeV which is the highest energy
level up to now. If one takes, for example, $n=2$, $m=3$ one has an average
radius of about 1.24fm. Also it is found that the average radii of \ $\Sigma
^{-}$\ are much smaller than those of the proton for levels with the same
quantum numbers $n$, $m$, $k$.

Now one can turn to $\Xi ^{-}$ which in terms of quarks is $dss$. Then it is
expected to have the same $k_{3}\approx 0.80$GeV/fm$^{2}$ (two of them)
above and a new $k$, which can be called $k_{ss}$. Using the ground state
radius of \ $\Xi ^{-}$ (0.66fm) one obtains $k_{ss}\approx 1.47$GeV/fm$^{2}$%
. For the excited states the average radius (in fm) is thus 
\begin{equation}
\mathcal{R}(n,m,k)=\left( \sqrt{\frac{0.31(n+1)\times 0.5(m+1)\times 0.5(k+1)%
}{1.47(0.8)^{2}}}\right) ^{1/3}
\end{equation}
which for the highest known excited state 2.55GeV ($n=4$, $m=k=0$) gives $%
\mathcal{R}(4,0,0)\approx 1.48$fm. Using the value $k_{ss}\approx 1.47$GeV/fm%
$^{2}$ the radius of the ground state of $\Omega $ is estimated to be about
0.58fm.

Putting together the above values the very important table below (Table 8)
is obtained for the constant $k$ (which is a sort of constant of
confinement) in terms of the pairs of interacting quarks.

\noindent The table shows that $k$ increases with the reduced mass of the
pair of interacting quarks. When the data are fitted to a polynomial up to
second order in the reduced mass of the pair of interacting quarks the
following polynomial is obtained 
\begin{equation}
k(\mu )=0.1188-1.7561\mu +28.6508\mu ^{2}.
\end{equation}
It is interesting that the coeficient of the last term is quite large and
thus the first derivative increases very rapidly with $\mu $. As more
massive quarks are considered the degree of the polynomial may increase but
just to have a lower bound one can calculate the value of $k$ for the
interaction between two top quarks. The above formula gives $k=7416$GeV/fm$%
^{2}$. If the above data are fitted to a polynomial with a higher degree,
for example, $k(\mu )=A+B\mu ^{2}+C\mu ^{3}$, the following values are
obtained: $A=-0.0268$, $B=23.3794$, and $C=0.2278$. Since the value of $C$
is small and $B$ is of the same order of 28.6508, the first polynomial (Eq.
26) is a good approximation. If it is used for obtaining the $k$ between
quarks $u$ and $c$ one has $k(uc)\approx 1.53$GeV/fm$^{2}$, $k(sc)\approx
3.7 $GeV/fm$^{2}$, $k(cc)\approx 19$GeV/fm$^{2}$. And then one has that the
radii of the ground states of the charmed baryons $\Lambda _{c}^{+}$, $%
\Sigma _{c}^{++}$, $\Sigma _{c}^{+}$ and $\Sigma _{c}^{0}$ are about 
\[
R_{c}\approx \left( \sqrt{\frac{0.31\times 0.31\times 1.7}{0.5(1.53)^{2}}}%
\right) ^{1/3}\approx 0.7\text{fm} 
\]

\noindent which is not so small due to the influence of the interaction
between the two $u$ quarks. As to $\Omega _{c}$ its ground state has a radius

\[
R_{ssc}\approx \left( \sqrt{\frac{0.5\times 0.5\times 1.7}{1.53(3.7)^{2}}}%
\right) ^{1/3}\approx 0.5\text{fm} 
\]

\noindent and the ground state of the $ccc$ baryon has the quite small
radius of just

\[
R_{ccc}\approx \sqrt{\frac{1.7}{19}}\approx 0.3\text{fm.} 
\]

\noindent Since the value of $k(cc)\approx 19$GeV/fm$^{2}$ was obtained by
means of an extrapolation the above figure of $R_{ccc}$ should be taken as a
crude approximation.

In the case of the ${ttt}$ baryon an even cruder number is gotten for its
radius because its value for $k$ is expected to be larger than the above
figure of $7416$GeV/fm$^{2},$\ but it is instructive anyway to calculate its
order of magnitude which in this case produces an upper bound for its
radius. Therefore one can say that the radius of the ground state of the ${%
ttt}$ system 
\[
R_{ttt}<\sqrt{\frac{174}{7416}}=0.15\text{fm.} 
\]

\noindent which is a very important number just because the top quark is the
most massive quark.

Since in this work the motion of the plane where quarks are sitting was not
taken into account conclusions can not be drawn on the shape of baryons
using the above figures.

\section{On the spin-orbit interaction}

We clearly notice that the splittings of some levels are caused by the
spin-orbit interaction. For example, consider the states $1.90S_{31}$ and $%
1.91P_{31}$ of $\Delta $ which differ by the values of $l=1$ and $l=0$,
respectively. Since we are assuming a harmonic potential and as the
spin-orbit term is proportional to $\frac{1}{r}\frac{dV}{dr}$ we can
approximately write

\begin{equation}
\Delta E_{SL}\approx C<\overrightarrow{S}.\overrightarrow{L}>=C\left[
j(j+1)-l(l+1)-s(s+1)\right]
\end{equation}
for $N$ and $\Delta $ baryons which have quarks with equal masses. Using for
the above case $j=1/2,s=1/2$ we find $C\approx 5$MeV which shows that the
influence of the spin-orbit interaction is small. Considering the levels $%
1.91P_{31}$ and $1.92P_{33}$ we find $C\approx 3.3$MeV which is of the same
order of the above $C$. The same holds in the case of the other baryons: for
example, consider the states $1.75S_{11}$ and $1.77P_{11}$ of $\Sigma $ or
the states $1.80S_{01}$ and $1.81P_{01}$ of $\Lambda .$ We see that there is
a small energy difference between these states.

\section{Conclusion}

The simple model presented above which considers that a baryon is composed
of three nonrelativistic quarks produces very importante results. First of
all it describes quite well almost all energy levels of baryons with the
proper assignment of the states as to parity and angular momentum. And the
calculation also yields some reasonable figures for the sizes of baryons and
for the important harmonic oscillator constant $k$ which is directly related
to confinement.

\pagebreak

\noindent {\Large References}

\noindent 1. S. Gasiorowicz and J.L.Rosner, Am. J. Phys., 49, 954(1981).

\noindent 2. N. Isgur and G. Karl, Phys. Rev. D18, 4187 (1978).

\noindent 3. S. Capstick and N. Isgur, \ Phys. Rev. D34, 2809 (1986).

\noindent 4. R.K. Bhaduri, B.K. Jennings and J.C. Waddington, Phys. Rev.
D29, 2051 (1984).

\noindent 5. M.V.N. Murthy, M. Dey, J. Dey and R.K. Bhaduri, Phys. Rev. D30,
152 (1984).

\noindent 6. M.V.N. Murthy, M. Brack, R.K. Bhaduri and B.K. Jennings, Z.
Phys. C29, 385 (1985).

\noindent 7. P. Stassat, Fl. Stancu and J.-M. Richard, nucl-th/9905015.

\noindent 8. A. Hosaka, H. Toki, and M. Tokayama, Mod. Phys. Lett A13, 1699
(1998).

\noindent 9. M.E. de Souza, Proceedings of the XIV Brazilian National
Meeting of the Physics of Particles and Fields, Caxambu, Minas Gerais,
Brazil, September 29-October 3, p. 331(1993).

\noindent 10. M.E. de Souza, in The Six Fundamental Forces of Nature,
Universidade Federal de Sergipe, S\~{a}o Crist\'{o}v\~{a}o, Sergipe, Brazil,
February 1994.

\noindent 11. M.E. de Souza, in Some Important Consequences of the Existence
of the Superstrong Interaction, Universidade Federal de Sergipe, S\~{a}o
Crist\'{o}v\~{a}o, Sergipe, Brazil, October, 1997.

\noindent 12. Particle Data Group, Phys. Rev. D54 (Part I), (1996).

\noindent 13. R. Shankar, in Principles of Quantum Mechanics, 2nd ed.,
Plenum Press, New York, 1994, pp 316-317. \ 

\noindent 14. B. Povh, hep-ph/9908233.

\noindent 15. Fayyazuddin and Riazuddin, in A modern Introduction to
Particle Physics, World Scientific, Singapore, 1992, pp 249-256.

\noindent 16. D. H. Perkins, in Introduction to High Energy Physics, 3rd
ed., Addison-Wesley Publishing Company Inc., Menlo Park, 1987, pp. 177-178.

\end{document}